\documentclass[10pt,twocolumn,final,conference,oneside]{IEEEtran}
\pdfoutput=1
\usepackage[top=25.4mm, left=19.1mm, bottom=19.1mm,right=19.1mm]{geometry}
\usepackage{cite}
\usepackage[T1]{fontenc}
\usepackage[latin1]{inputenc}
\usepackage{booktabs} 
\usepackage{textcomp} 
\usepackage{amssymb}
\usepackage[cmex10]{amsmath}
\usepackage{upgreek} 
\usepackage{subfigure}

\renewcommand{\eqref}[1]{(\ref{#1})}

\DeclareMathOperator*{\argmin}{arg\;min}

\DeclareMathOperator*{\sign}{sign}

\newcommand{\setC}{\ensuremath{\mathcal{C}}}
\newcommand{\setD}{\ensuremath{\mathcal{D}}}

\newcommand{\setN}{\ensuremath{\mathcal{N}}}

\newcommand{\setQ}{\ensuremath{\mathcal{Q}}}

\newcommand{\bms}{\ensuremath{\mathbf{s}}}

\newcommand{\bmsbar}{\ensuremath{\bar{\bms}}}

\ifCLASSINFOpdf
   \usepackage[pdftex]{graphicx}
   \graphicspath{{./fig/pdf/}{./fig/jpeg/}}
   \DeclareGraphicsExtensions{.pdf,.jpeg,.png}
\else
\fi

\IEEEoverridecommandlockouts 

\usepackage{chngcntr}
\counterwithout{paragraph}{subsubsection}
\counterwithin*{paragraph}{section}

\newcommand{\figspace}{\vspace{-0.2cm}}
\newcommand{\capspace}{\vspace{-0.2cm}}

\begin{document}

\title{Data Mapping for Unreliable Memories}


\author{
\IEEEauthorblockN{Christoph~Roth$^\text{1}$,
  Christian~Benkeser$^\text{1}$,
  Christoph~Studer$^\text{2}$,
  Georgios~Karakonstantis$^\text{3}$,
  and Andreas~Burg$^\text{3}$}
\vspace{0.2cm} \\
\IEEEauthorblockA{$^\text{1}$Dept.\ of Information Technology
  and Electrical Engineering, ETH Z\"urich, 8092 Z\"urich,
  Switzerland}
\IEEEauthorblockA{$^\text{2}$Dept.\ of Electrical and
  Computer Engineering, Rice University,
  Houston, 77005 TX, USA}
\IEEEauthorblockA{$^\text{3}$School of Engineering, EPF
  Lausanne, 1015 Lausanne, Switzerland}
\begin{minipage}[t]{1\textwidth}
  \centering
  e-mail: \{rothc,\,benkeser\}@iis.ee.ethz.ch; studer@rice.edu; \{georgios.karakonstantis,\,andreas.burg\}@epfl.ch
\end{minipage}
\thanks{The work at EPFL was supported by the Swiss National Science Foundation (SNSF) under grant PP002-119052. The work of C.~Studer was supported in part by the SNSF grant PA00P2-134155.}
}


\maketitle

\begin{abstract}
Future digital signal processing (DSP) systems must provide robustness
on algorithm and application level to the presence of reliability issues that come along with corresponding implementations in modern semiconductor process technologies. 
In this paper, we address this issue by investigating the impact of unreliable memories on general DSP systems.
In particular, we propose a novel framework to characterize the effects of unreliable memories, which enables us to devise novel methods to mitigate the associated performance loss.
We propose to deploy specifically designed data representations, which have the capability of substantially improving the system reliability compared to that realized by conventional data representations used in digital integrated circuits, such as 2's-complement or sign-magnitude number formats.
To demonstrate the efficacy of the proposed framework, we analyze the impact of unreliable memories on coded communication systems, and we show that the deployment of optimized data representations substantially improves the error-rate performance of such systems.

%
%
\end{abstract}


\section{Introduction}
\label{sec:introduction}

The continuous shrinkage of semiconductor devices during recent years
has led to the enormous success of digital signal processing~(DSP)
systems by enabling the realization of ever more sophisticated and
powerful implementations, which have become ubiquitous in our daily lives. Such
an evolution was---up to now---relying on the assumption that the
underlying hardware is able to perform computations and store data in
a 100\% reliable manner. However, it is now becoming apparent that
such a trend may come to an end due to the increasing effect of semiconductor-process
variability as well as reliability issues that threaten the correct circuit
functionality, especially for CMOS technology nodes beyond 45\,nm.
Specifically, the small size of semiconductor devices in
combination with delicate fabrication processes lead to a wide
variability of transistor characteristics inside and among fabricated
dies that may lead to failures. In order to maintain an acceptable fabrication yield under these
conditions, the inclusion of design guard-bands and the design
for the worst case has become a necessity~\cite{rabaey2009low,
  bhunia2010low, Borkar2004}. Unfortunately, these precautions are
costly in terms of the associated silicon area- and power overhead and, thus, render
the design of cost-effective and power-efficient DSP systems
significantly challenging. In addition, the application of effective
low-power techniques, such as aggressive voltage-frequency scaling, makes the
fabricated systems even more sensitive to process variations, which further
diminishes their use in practice~\mbox{\!\!\cite{Chishti2009, Wilkerson2008}}.

Memories are particularly susceptible to process variations as their
operation is highly dependent on the characteristics between
neighboring transistors on the fabricated die. In order to ensure 
100\% reliable operation, costly error-correction coding schemes or
novel bit-cell topologies are required to tackle the high failure
probability of traditional bit-cells, which can lead to more than 30\%
area overhead~\cite{Zhou2010, Emre2010}. In addition, these techniques
come with a significant overhead in terms of energy efficiency, which is rather unacceptable
considering the fact that the share of memories in modern digital
systems is continuously increasing~\cite{Tensilica2009}. In many cases, such error-correction mechanisms
evidently waste resources since they serve merely as precaution
against worst-case conditions, which only occur in a
fraction of the fabricated circuits and memory cells. 
For some 
applications, such as wireless communication or multimedia, memories are already dominating both silicon
area and power consumption.

Consequently, realizing cost-effective and energy-efficient DSP systems in
the near future requires a paradigm shift from the assumption of 100\%
reliable operation to fault-tolerant DSP systems that are
robust against implementation on unreliable silicon~\cite{gosh10}. Interestingly,
various applications and algorithms~(e.g.,~for wireless communication systems
or video processing) are inherently fault tolerant as they are
naturally able to deal with stochastic data already corrupted by noise
and interference. It has been observed that this inherent
fault-tolerance can be exploited to allow for a certain amount of errors
induced by unreliable silicon components without degrading the system
performance noticeably, provided that the corresponding algorithms and
system architectures are designed to take such hardware errors into
account~\cite{novak10,Hussien2010, Karakonstantis2010, Karakonstantis2012}.


\subsection{Contributions}

\begin{figure*}[tb]
  \centering
  \subfigure[General DSP system with an unreliable memory.]{\includegraphics[width=1.45\columnwidth]{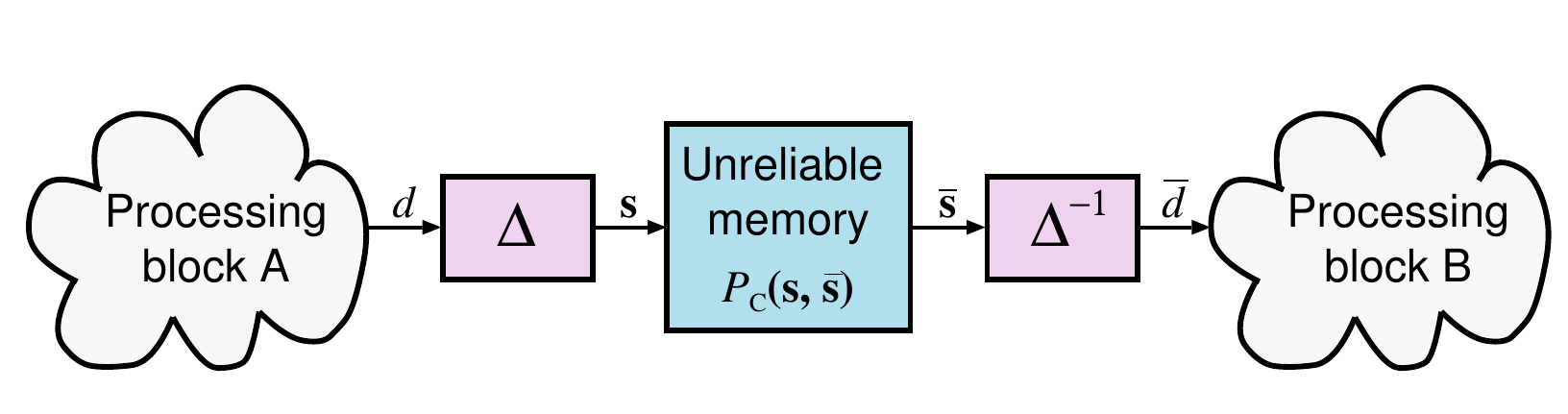}}
  \hfill
  \subfigure[Data recoding for unreliable memories.]{\includegraphics[width=1.45\columnwidth]{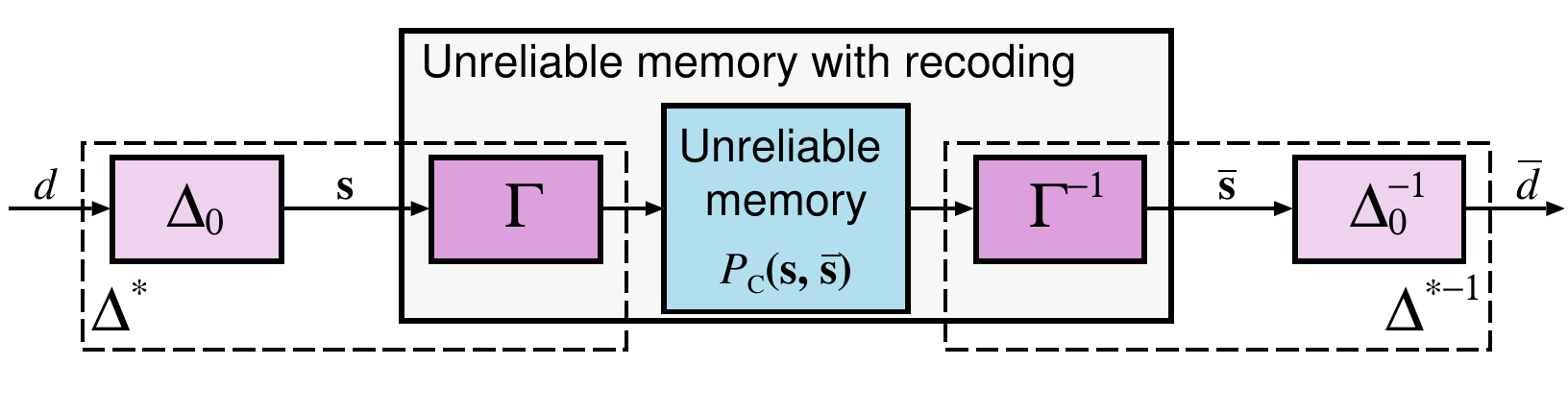}}
  \caption{Data mapping and recoding in DSP systems. The data mapping function $\Delta$ maps the digital data symbols to binary-valued labels of length $N$ that are stored in the unreliable data memory. The bijective recoding function $\Gamma$ converts a conventional mapping into an arbitrary custom mapping.}
  \label{fig:dataMapping}
  \capspace
\end{figure*}

By arguing that memories in modern DSP systems are particularly prone to
unreliable operation, we investigate the impact of defects in memories
on the performance of DSP systems. Specifically, we propose to reduce this impact by choosing data
representations that are different from the ones typically
employed in digital integrated circuits. Our contributions are summarized as follows:
%
%
\begin{itemize}
\item We present a novel framework to analyze the impact of data
representations on their performance in unreliable DSP systems.
\item We introduce the ``stuck-at-channel'' as a realistic model for bit-cells in unreliable memories.
\item We show that the choice of the data representation has a significant impact on
the system performance in presence of unreliable memories.
\item We formulate the problem of finding favorable data representations
as an optimization problem based on the memory input-data distribution
and an application-specific cost function.
\end{itemize}
As a proof of concept, we apply the proposed framework to a
coded communication system and show that optimized data representations
significantly increase the robustness against defects in memories compared to that of data representations commonly used in digital integrated circuits.

\subsection{Paper Outline}

The remainder of this paper is organized as
follows. In Sec.~\ref{sec:dataMapping}, we introduce the framework and discuss the 
impact of the choice of different (non-redundant) binary-valued data representations  
in the context of unreliable memories. We also show how different data representations 
can be optimized for improving robustness against unreliable memories.
In Sec.~\ref{sec:applCommSys}, we consider a coded communication system as a 
specific application example, and we provide simulation results to quantify the impact of optimized data representations on two relevant coding schemes. We 
conclude in Sec.~\ref{sec:conclusion}.



\section{Optimized Data Representations for Unreliable Memories}
\label{sec:dataMapping}


%

We consider an unreliable memory as depicted in Fig.~\ref{fig:dataMapping}a), which is surrounded by two processing blocks in a general DSP
system. The discrete input
data~$d\in \setD$ with $\setD = \{d_1,\ldots,d_K\}$ is assumed to be
distributed according to the probability mass function \mbox{$P_d(d_k)
=\mathrm{Pr}(d=d_k)$}. In order to store data in the unreliable memory, each
symbol~$d_k$ is mapped to a vector $\bms_k$ with binary-valued entries and  
dimension $N=\log_2\!K$ bit. We refer to $\bms_k$ as the (binary-valued) label 
of $d_k$. In what follows, we make frequent use of the bijective 
mapping function $\Delta$ and its inverse $\Delta^{-1}$, which implement $\bms_k =
\Delta(d_k)$ and \mbox{$d_k = \Delta^{-1}(\bms_k)$}; the mapping function defines the (binary-valued) 
\emph{data-representation}.

Commonly used data representations in digital systems are the \mbox{2's complement}~(2C)
or the sign-magnitude~(SM) number formats. Both data representations allow for the 
efficient implementation of basic arithmetic
operations in digital integrated circuits~\cite{kaeslin08}. As it will be shown next, the choice of
the data representation has a significant impact on the robustness of
systems containing unreliable memories. Furthermore, we will show that
2C or SM are not necessarily good choices in many
practical application scenarios.

\subsection{Probabilistic Models for Unreliable Memories}
In the remainder of the paper, we focus on unreliable memories in which the stored 
labels are subject to \emph{probabilistic} errors.
In particular, we model the process of storing the input data values~$d\in \setD$ (physically represented by its label) 
as an i.i.d.\ non-binary probabilistic
channel that maps input labels $\bms$ to output labels
$\bmsbar$ according to a label cross-over probability mass function 
$P_{\mathrm{C}}(\bms_k, \bmsbar_{k'}) =
\mathrm{Pr}(\bmsbar=\bmsbar_{k'} | \bms=\bms_k)$
(see~Fig.~\ref{fig:dataMapping}a).
This label cross-over probability depends on the employed  model for physical
memory bit-cell errors, which are not necessarily uncorrelated within the 
bits belonging to the same symbol.\footnote{In reality, errors could also be correlated 
among the bits of different symbols. For the sake of simplicity of exposition, however, our focus is on this simplified model.}

\subsubsection{Binary-symmetric channel (BSC)}

A straightforward model for errors in the individual bit-cells of 
unreliable memories is the BSC~\cite{novak10,Hussien2010}. In this case, the non-binary channel $\bms_k\rightarrow\bmsbar_{k'}$ is described 
by parallel and independent BSCs for each bit-cell with bit-flip probability~$\varepsilon$. 
The associated label cross-over probabilities thus correspond to~\cite{novak10}
\begin{align*}
    P_{\text{C,BSC}}(\bms_k, \bmsbar_{k'}) = \varepsilon^{d_H(\bms_k,\bmsbar_{k'})} (1-\varepsilon)^{N-d_H(\bms_k,\bmsbar_{k'})},
\end{align*}
where $d_H(\bms_k, \bmsbar_{k'})$ denotes the Hamming distance
between the two labels $\bms_k$ and $\bmsbar_{k'}$.

\begin{figure}[t]
  \centering
  \includegraphics[width=0.9\columnwidth]{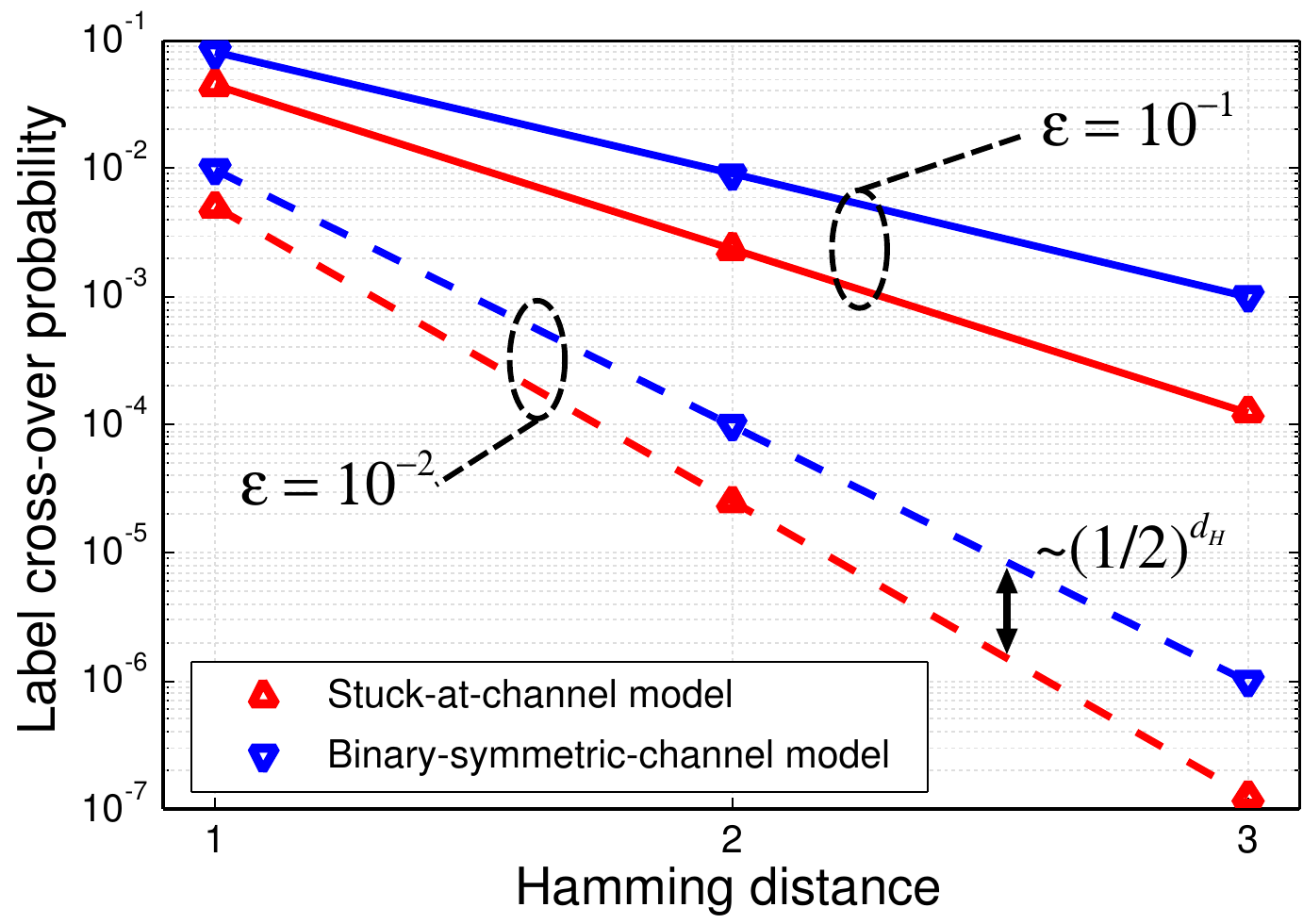}
  \figspace
  \caption{Label cross-over probability comparison of the 
    binary-symmetric-channel (BSC) and stuck-at-channel (SAC) fault-models.}
  \label{fig:bscvsStuckat}
  \capspace
\end{figure}

\subsubsection{Stuck-at channel}
Unfortunately, the BSC fault model is not particularly relevant when
considering embedded memories with very-large scale integration~(VLSI) in
nanometer silicon processes that are affected by process
variations~\cite{dekker06}. 
To arrive at a more realistic model of the true physical effects, we propose a new channel model that is more 
representative for the typically applied fault model for 
VLSI memories. We refer to this new channel model as the \emph{stuck-at channel}~(SAC). 
The fault model on which the SAC is based upon assumes that 
each bit-cell fails independently with a bit-cell error probability~$\varepsilon$. However, in contrast to the BSC model, 
the stuck-at-channel model
further assumes that a faulty bit cell is either stuck-at-0 or
stuck-at-1 with equal probability. Physically, this effect refers to a node in the circuit being either shortened to power 
or ground, or to the inability of a circuit or transistor to pull a node sufficiently strong to either side of the rails. 
Consequently, the label cross-over probabilities for the SAC fault model correspond~to 
\begin{align}
\label{eq:coprob}
    P_{\text{C,SAC}}(\bms_k, \bmsbar_{k'}) &= 
    \sum_{\ell=0}^{N-d_H(\bms_k,\bmsbar_{k'})}{N-d_H(\bms_k,\bmsbar_{k'})
      \choose \ell} \times \notag \\
    &\!\!\!\!\left( \frac{\varepsilon}{2}
    \right)^{d_H(\bms_k,\bmsbar_{k'})+\ell}
    \left( 1-\varepsilon
    \right)^{N-d_H(\bms_k,\bmsbar_{k'})-\ell}.
\end{align}

\subsubsection{Comparison between BSC and SAC}

In Fig.~\ref{fig:bscvsStuckat}, we compare the BSC and SAC fault models by
plotting their label cross-over probability $P_{\text{C,BSC}}(\bms_k,
\bmsbar_{k'})$ and $P_{\text{C,SAC}}(\bms_k,
\bmsbar_{k'})$ depending on the Hamming distance $d_H(\bms_k, \bmsbar_{k'})$
for $N=3$, and $\varepsilon =10^{-1}$ and $\varepsilon = 10^{-2}$.
We can observe that the BSC fault model is rather pessimistic compared to the SAC fault model, especially for
large Hamming distances. 
In the remainder of this paper, we use exclusively study the SAC fault model, since it reflects the behavior of 
practical implementations more accurately. We emphasize, however, that the general discussion and observations remains valid for other fault models.

\subsection{Optimized Data Representations}
In order to minimize the amount of errors introduced by unreliable memories, without adding redundancy 
to the data representation (e.g., by means of coding), we propose to use specifically optimized data representations. 
The choice of such non-redundant representations is driven by the desire to minimize the associated hardware overhead and to sustain as much system performance as possible, even if data is not stored 100\% reliably. 

\subsubsection{Optimal data representations}
We start by characterizing the impact of the unreliable memory on the system with an
application-specific cost function $\setC(\Delta)$ that depends on the data-mapping function $\Delta$. We then 
choose an \emph{optimized data representation} defined by the mapping $\Delta^*$, which minimizes the  cost function $\setC$ for the specified
memory input-data distribution $P_d$ and for the label cross-over
probability $P_{\text{C,SAC}}$ according to 
\begin{align} \label{eq:opt}
    \Delta^*= \argmin_{\Delta'} \, \setC(\Delta').
\end{align}

We emphasize that the number of choices for representing~$2^N$ data symbols is $(2^N)!$. While for 
$N=3$ bit an exhaustive evaluation of the 40'320 possible candidate mappings is still feasible, $N=4$ requires already 
the consideration of $2.09\cdot10^{13}$ candidates. Hence, for $N>3$ one must resort to sophisticated algorithms  to identify good data representations. 
In the following, we  limit our studies to $N=3$; the development of optimization algorithms suitable for $N>3$ is part of on-going work.


\subsubsection{Example cost function}
\begin{figure}[t]
  \centering
  \includegraphics[width=0.95\columnwidth]{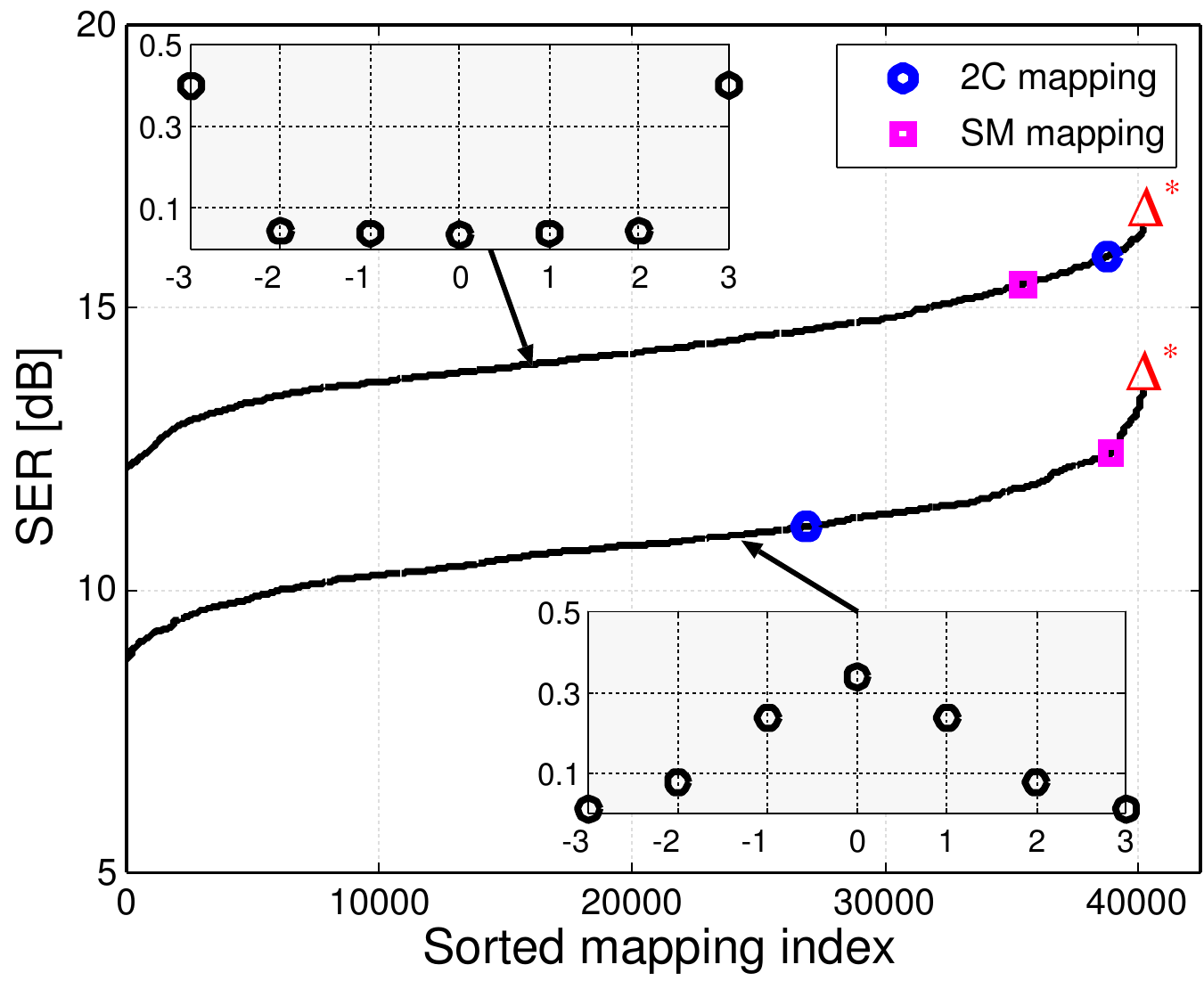}
  \figspace
  \caption{Impact of data representation on the signal-to-mean-squared-error ratio (SER) assuming two example memory
    input-data distributions. The  
    bit-cell error probability is \mbox{$\varepsilon=10^{-2}$}.}
  \label{fig:effMappingSMER}
  \capspace
\end{figure}

Consider the mean
squared error~(MSE) between the memory input $d$ and the corrupted output $\bar{d}$ as an example for a cost function $\setC$.  
With this, we can measure the amount of errors induced by the SAC fault model
depending on the mapping function~$\Delta$ as follows:
\begin{align}
\label{eq:cmse}
\setC_{\text{MSE}}(\Delta) & =  
\sum_{k=1}^{K} P_d(\Delta^{-1}(\bms_k)) \times \notag \\ & \sum_{k'=1}^K (\Delta^{-1}(\bms_k)\!-\!\Delta^{-1}(\bmsbar_{k'}))^2 P_{\text{C,SAC}}(\bms_k, \bmsbar_{k'})).
\end{align}

To illustrate the impact of the mapping function $\Delta$ on the MSE we consider 
an input symbol alphabet of 8 integers $\setD = \{-4, -3, -2,\ldots, 3\}$ and two different 
distributions\footnote{To maintain symmetry, we set $P_d(-4) = 0$.} as shown in Fig.~\ref{fig:effMappingSMER}. 
In this example, the bit-cell error probability of the unreliable memory is set to~$\varepsilon =10^{-2}$. 
In order to properly account for differences in the signal power for the two considered distributions, we show 
the signal-to-mean-squared-error ratio~(SER) defined as $\text{SER}=\mathbb{E}[d^2]/\text{MSE}$ 
for all possible mappings in Fig.~\ref{fig:effMappingSMER}. 
We observe that the choice of the
data mapping function $\Delta$ has, indeed, a significant impact on the quality of the 
signal at the output of the unreliable memory. 
Specifically, the different data representations cover an SER range of roughly 4.5\,dB with the commonly used 
2C and SM representations clearly outperformed by the corresponding optimal mapping functions $\Delta^*$.



\subsection{Data Recoding}
As observed in the example above, the choice of the data representation has a significant impact on the MSE incurred by unreliable memories and, hence, must be optimized in order to achieve maximum robustness. 
However, the use of optimized custom data representations is very impractical for digital systems as they usually rely on specific data representations that enable the efficient implementation of arithmetic operations in hardware. Hence, we
 propose \emph{data recoding} as an approach to render the
issue of custom data representations for memories \emph{transparent} to the
surrounding system. With this approach, we let the system employ a
fixed (and hardware-friendly) data mapping $\Delta$ but add a recoding function
$\Gamma$ to the memory input as depicted in
Fig.~\ref{fig:dataMapping}b). The recoding function relabels the memory input-labels
such that $(\Gamma \circ \Delta) = \Delta^*$.  At the
memory output, the inverse recoding function $\Gamma^{-1}$ is
applied to recover the data representation employed by the surrounding
system. Note that the recoding function can be implemented by simple
look-up tables. Therefore, data recoding can be considered to be a low-complexity 
technique to increase the robustness of DSP systems containing
unreliable memories, while leaving the surrounding processing blocks untouched. 


\section{Application to Communication Systems}
\label{sec:applCommSys}

As an application example of the proposed framework, we now study the impact of data representations on the performance of a coded digital communication receiver containing unreliable memories. 
First, we investigate the general impact of data representations on the achievable communication rates and then, we analyze the associated error-rate performance for repetition coding and convolutional codes.

\begin{figure*}[t]
  \centering
  \includegraphics[width=0.95\textwidth]{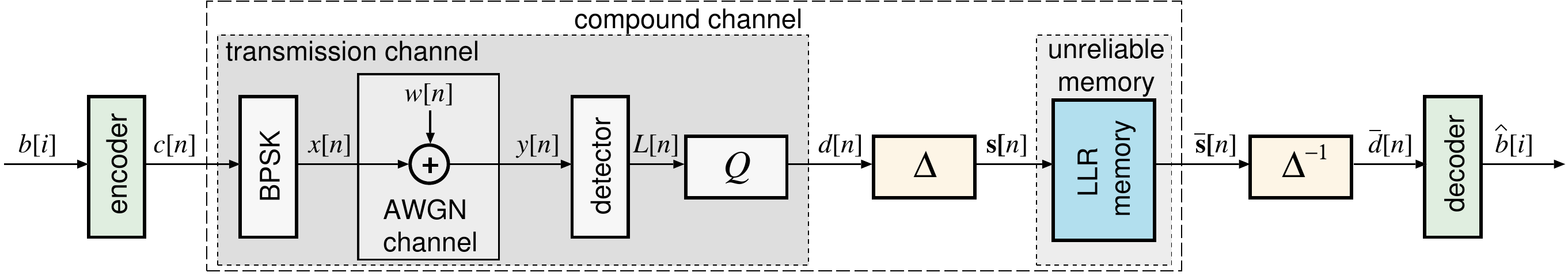}
  \figspace
  \caption{Digital communication system employing BPSK 
    transmission over an AWGN channel. The receiver consists of a
    soft-output detector, an LLR quantization block, a data mapping
    stage $\Delta$, and unreliable LLR memory. At the memory
    output, the stored labels are remapped to symbols using the inverse data
    mapping function $\Delta^{-1}$ and then passed to a soft-input decoder.}
  \label{fig:commSysModel}
  \capspace
\end{figure*}

\subsection{System Model}

\subsubsection{Transmitter and AWGN channel}
We consider the communication system introduced in \cite{novak10} and depicted in Fig.~\ref{fig:commSysModel}. 
A sequence of information bits $b[i]$, $i=1,\ldots,N_M$, is encoded into a sequence of
coded bits $c[n]$, \mbox{$n=1,\ldots,N_C$}, using a rate $N_M/N_C$ forward error-correction code. The coded bits are then
 mapped to binary phase shift keying (BPSK) symbols $x[n] \in \{+1,-1\}$ that are
transmitted over an AWGN channel, modeled as \mbox{$y[n]=c[n]+w[n]$}, where~$y[n]$ denotes the received signal and $w[n] \sim \setN(0, \sigma^2)$. At the receiver, a 
soft-output detector computes log-likelihood
ratio~(LLR) values for each coded bit $c[n]$ based on the received signal $y[n]$ and the noise variance $\sigma^2$ according to
$L[n] = {2y[n]}/{\sigma^2}$ \cite{Hagenauer96}. 

\subsubsection{Receiver with LLR quantizer}
In the receiver, the LLR values are passed through
a uniform $N$-bit scalar quantizer~$\setQ$. We assume a scalar quantization scheme with $F$ fractional
bits and $I=N-1-F$ integer bits such that \mbox{$\setD =
\{-2^{I-F},-2^{I-F}+2^{-F}, \ldots, 2^{I-F}-2^{-F} \}$}. The effect of the quantizer is defined as~\cite{chen05}
\begin{align}
    \label{eq:quant}
    d[n] = 2^F \sign(L[n]) \min \!\left( \left \lfloor
             \delta\frac{|L[n]|}{2^F}+\frac{1}{2} \right \rfloor, 2^I-1 \right)\!,
\end{align}
where $\lfloor\cdot\rfloor$ denotes rounding towards $-\infty$.
Note that~(\ref{eq:quant}) enforces a symmetric output distribution of
the quantized LLRs with $P_d(d_1) = 0$. 
Also note that due to $P_d(d_1) = 0$, the label $\bms_1$ becomes a redundant label that
is not used by the system. In order to cope with the possibility that this
label appears at the output of an unreliable memory, we define
$\Delta^{-1}(\bms_1) = d_2$.
We note that the quantization scheme in~\eqref{eq:quant} allows for a
scaling of the input LLRs by the parameter $\delta>0$, which allows one to adjust the
quantization range to the application at hand (see, e.g., \cite{chen05,roth10}).

\subsubsection{Unreliable memory}
In the communication system shown in Fig.~\ref{fig:commSysModel}, the quantized LLR values are then mapped to binary-valued labels $\bms$ using the data mapping function~$\Delta$ and stored in an unreliable LLR memory with label
cross-over probabilities $P_\text{C,SAC}(\bms,\bmsbar)$ following the stuck-at-channel model.
In practice, an unreliable memory could be used for data (de-)interleaving or as a large buffer that stores the LLR values of several data (re-)transmissions in modern wireless communication systems employing hybrid-ARQ (automatic repeat-request), such as 3GPP-HSPA~\cite{3GPP_TS25222}.


\subsection{Impact of Data Representations on Mutual Information}

\subsubsection{Compound channel}

As depicted in Fig.~\ref{fig:commSysModel}, the combination of the
transmission channel and the unreliable memory forms a (memory-less) compound
channel with binary-valued channel inputs $c[n]$ and binary-valued output label vector~$\bmsbar[n]$~\cite{novak10}. The transmission channel is characterized by the
conditional probabilities of observing the channel output $d[n]=d_k$ given the  transmitted 
coded bit is $c[n]=b$, i.e., 
\begin{align}
    \label{eq:crossoverTC}
    P_{d|c}(d_k |b) = \mathrm{Pr}(d[n] = d_k | c[n]=b).
\end{align}
The compound channel, on the other hand, is characterized by the
conditional probabilities of observing the label $\bmsbar[n]$ at the
output of the unreliable LLR memory given that the coded bit $c[n]=b$ was
transmitted. Using~(\ref{eq:crossoverTC}) and the label cross-over
probability of the unreliable LLR memory modeled by~$P_\text{C,SAC}$, these probabilities correspond to 
\begin{align}
    \label{eq:crossoverCTC}
    P_{\bmsbar | c}(\bmsbar_k | b) =\sum_{k'=1}^K P_{d|c}(\Delta^{-1}(\bms_{k'})|b) P_\mathrm{C,SAC}(\bms_{k'}, \bmsbar_k),
\end{align}
which enables us to calculate the mutual information $I(c;\bmsbar)$ of the compound
channel as follows~\cite{cover1991}:
\begin{align}
    \label{eq:mutInf}
    I(c;\bmsbar) = & \frac{1}{2} \sum_{b\in\{0,1\}} \sum_{k=1}^{K} P_{\bmsbar |
      c}(\bmsbar_k | b) \times \notag \\ 
      & \qquad \log_2\!  \frac{2P_{\bmsbar | c}(\bmsbar_k | b)}{P_{\bmsbar | c}(\bmsbar_k | 0) + P_{\bmsbar | c}(\bmsbar_k | 1)}.
\end{align}

\subsubsection{Simulation results and discussion}

\begin{figure}[t]
  \centering
  \includegraphics[width=0.95\columnwidth]{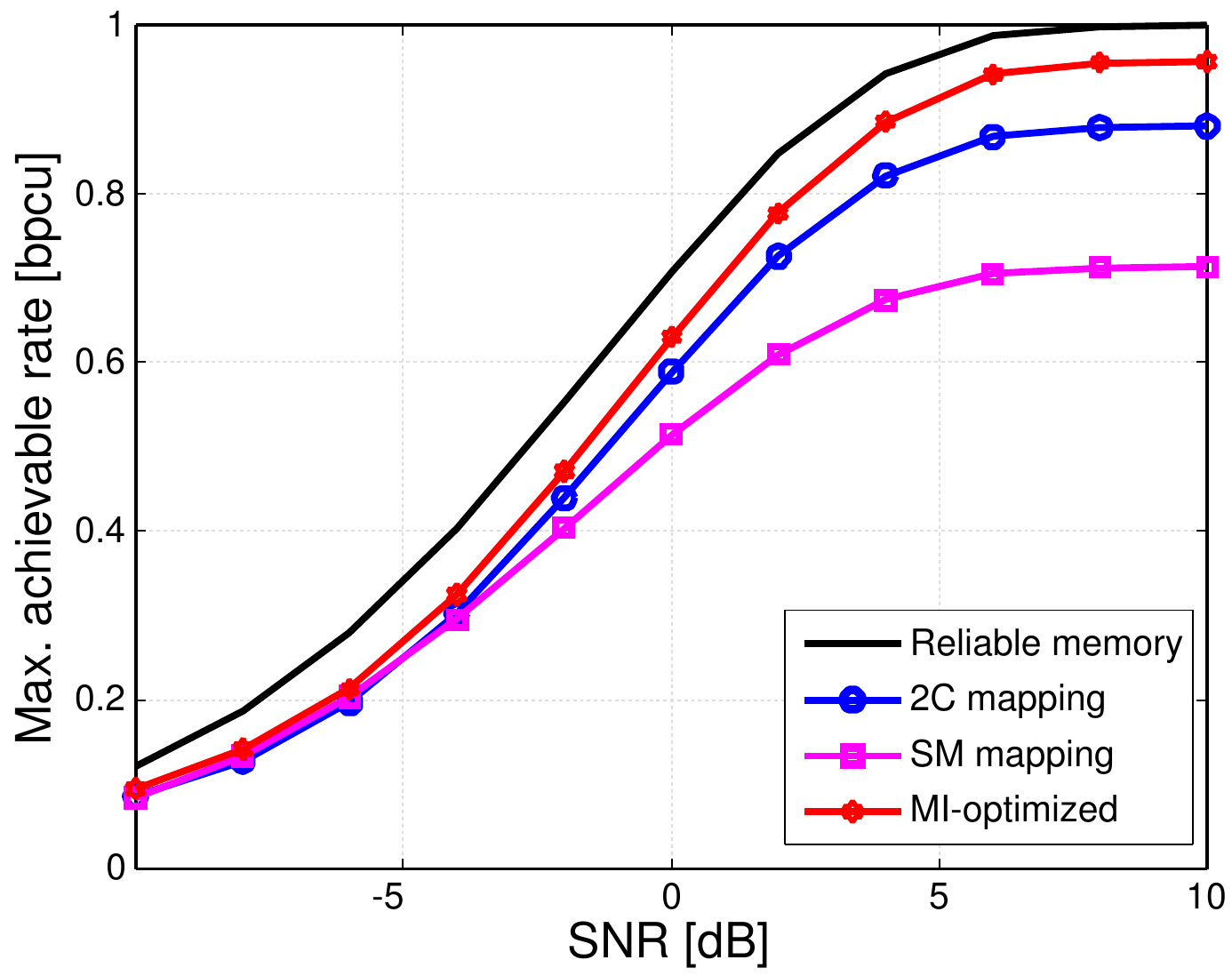}
  \figspace
  \caption{Maximum achievable rates of BPSK modulation over a compound channel consisting of an AWGN channel and unreliable LLR memory with bit-cell error probability $\varepsilon=10^{-1}$ assuming linear quantization (with parameters $N=3$, $I=2$, $F=0$, and $\delta=1$) and for different data representations.}
  \label{fig:achievRates}
  \capspace
\end{figure}

We now study the impact of data representations on the achievable
rate of the considered system, assuming a fixed LLR quantizer
with $N=3$ bit. To this end, we numerically determine
$P_{\bmsbar | c}(\bmsbar_k | b) $ with the aid of Monte-Carlo simulations for different AWGN-channel signal-to-noise
ratios~(SNR) and a memory bit-cell error probability of $\varepsilon=10^{-1}$. The mutual
information~(\ref{eq:mutInf}) is evaluated for different data mapping functions.
Fig.~\ref{fig:achievRates} shows the achievable rates for different
data representations. As a reference, we show the achievable rate
for a reliable LLR memory (with $\varepsilon=0$).
Furthermore,  we choose
\mbox{$\setC_{\mathrm{MI}} = -I(c;\bmsbar)$} as the cost function and perform an 
exhaustive search over all possible mappings for each SNR point (note that a different mapping may be required for each SNR).
The performance of the mappings delivering the highest achievable rates for each SNR level are shown in 
Fig.~\ref{fig:achievRates} (with label `MI-optimized'). 

We can observe
that, in contrast to the MI-optimized data mapping, the two common number representations
in digital circuits, namely 2C and SM, significantly limit the achievable rate, while 2C clearly
outperforms SM. 
The fact that SM is outperformed by 2C for high SNRs can be explained as follows:
In the high-SNR regime, the LLR distribution $P_{d}$ becomes
bimodal with the
most-positive and most-negative quantization bin~(i.e., $d_2$ and
$d_K$)  having the highest probability. In this case, a bit-flip of the sign bit for
the SM representation causes a large error, i.e., it maps $d_K$ to
$d_2$ and vice-versa with high probability.
Furthermore, both number representations incur a significant rate loss at low SNR values compared to the data mappings that maximize the mutual information. For example, at a rate of 0.5\,bit per channel
use~(bpcu) this loss is roughly 0.5\,dB and 1\,dB for 2C and SM, respectively.

\subsection{Optimized Data Mapping for Repetition Coding}

We next show that the chosen data representation has a significant impact on the bit error-rate (BER) performance. 
To this end, we assume that the communication system shown in
Fig.~\ref{fig:commSysModel} employs a rate-1/2 repetition code.
Repetition coding is the basic model for
hybrid-ARQ, which is a key feature in many modern wireless
communications systems, such as 3GPP-HSPA~\cite{3GPP_TS25222}, which specifies a rapid
retransmission of erroneously received data blocks. 
For systems employing hybrid-ARQ, the LLR memory in Fig.~\ref{fig:commSysModel}
serves as a buffer that stores the LLRs of the individual 
(re-)transmissions. In practical systems, this buffer can be very
large (i.e., storing the LLRs of multiple transmitted code blocks) and, hence, corresponding silicon realizations would benefit from high-density memories that are prone to unreliable operation. 

\subsubsection{System model}

In the example investigated here, we consider only two uncoded transmissions (i.e., \mbox{$N_M = N_C$}) and assume that
the quantized LLRs $d[n]$ of the first transmission are buffered in the unreliable
LLR memory and then combined with the quantized LLR values~$d'[n]$ from the second
transmission. The combined LLR values $d^\text{c}[n] = \bar{d}[n] + d'[n]$ are
then sliced to 0 or 1 by the decoder in Fig.~\ref{fig:commSysModel} depending on their sign.

\subsubsection{Cost function for repetition coding}
In order to minimize the impact of the unreliable LLR memory on
the system's error-rate performance, we are interested in a cost function for repetition coding that enables us to identify a  suitable data mapping function. Since the BER $P_e$ determines the performance of the system,
we set $\setC_{\text{REP}} = P_e$. 
For repetition coding, this cost function can be established analytically as the
sum of the probability that the combined LLR $d^\mathrm{c}[n]$ is smaller than zero having
sent $c=0$ and the probability that the $d^\mathrm{c}[n]$ is greater or equal
than 0 having sent $c=1$, i.e., 
\begin{align} \label{eq:peRep}
 P_e = & \frac{1}{2} \mathrm{Pr}(d^\mathrm{c}[n] < 0 | c=0) + \frac{1}{2}\mathrm{Pr}(d^\mathrm{c}[n] \geq 0 | c=1).
\end{align}

\subsubsection{Simulation results and discussion}
\label{sec:repCoding}
\begin{figure}
\centering
\includegraphics[width=0.95\columnwidth]{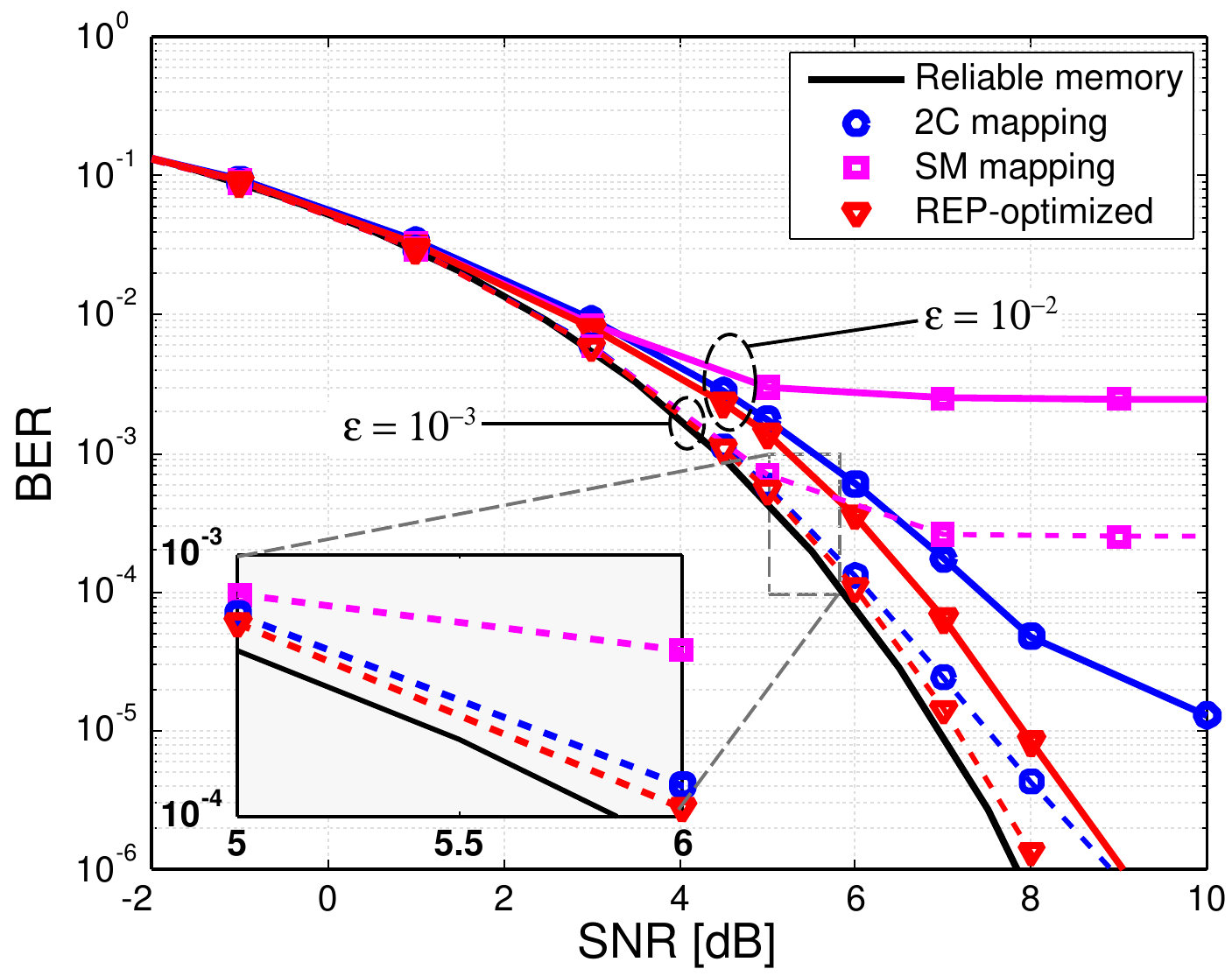}
\figspace
\caption{BER-performance comparison of repetition coding for different
  data representations. Solid curves represent the results
  for $\varepsilon=10^{-2}$, dashed curves the results for $\varepsilon =
  10^{-3}$. The quantizer parameters are $N=3$, $I=2$, $F=0$, and $\delta=0.2$.}
\label{fig:berRep}
\end{figure}
In Fig.~\ref{fig:berRep} we compare the BER performance of the repetition-coding system
for different data representations assuming two different bit-cell error
probabilities, i.e.,  $\varepsilon = 10^{-2}$~(solid curves) and
\mbox{$\varepsilon = 10^{-3}$}~(dashed curves). We set
$N=3$, which allows us to optimize $\setC_{\text{REP}}$
in~(\ref{eq:peRep}) exhaustively over all possible data
mappings for each SNR operating point. This approach
yields the optimal data mapping in terms of the system's BER performance (labeled `REP-optimized' in Fig.~\ref{fig:berRep}).

We see from Fig.~\ref{fig:berRep} that the data representation not only
affects the achievable rates~(cf.~Fig.~\ref{fig:achievRates}) but also the BER-performance
of the system employing a particular error-correction scheme. In this
particular  repetition-coding example for $\varepsilon=10^{-2}$, the commonly deployed data representations exhibit a high error-floor and suffer from severe SNR penalties; again, 2C outperforms SM. The optimized mapping, however, yields significantly better BER performance.
For higher bit-cell reliability, e.g., $\varepsilon = 10^{-3}$, the BER-performance loss of 2C and SM
to the ones optimized for $C_{\text{REP}}$ is still pronounced. While SM exhibits an unacceptably high error floor, the 2C data representation loses roughly 0.5\,dB and 1\,dB at a BER of 10$^{-5}$ and 10$^{-6}$, respectively, compared to the mappings optimized for repetition coding.

\subsection{Optimized Data Mapping for Convolutional Coding}
\label{sec:convCoding}
\begin{figure}
\centering
\includegraphics[width=0.948\columnwidth]{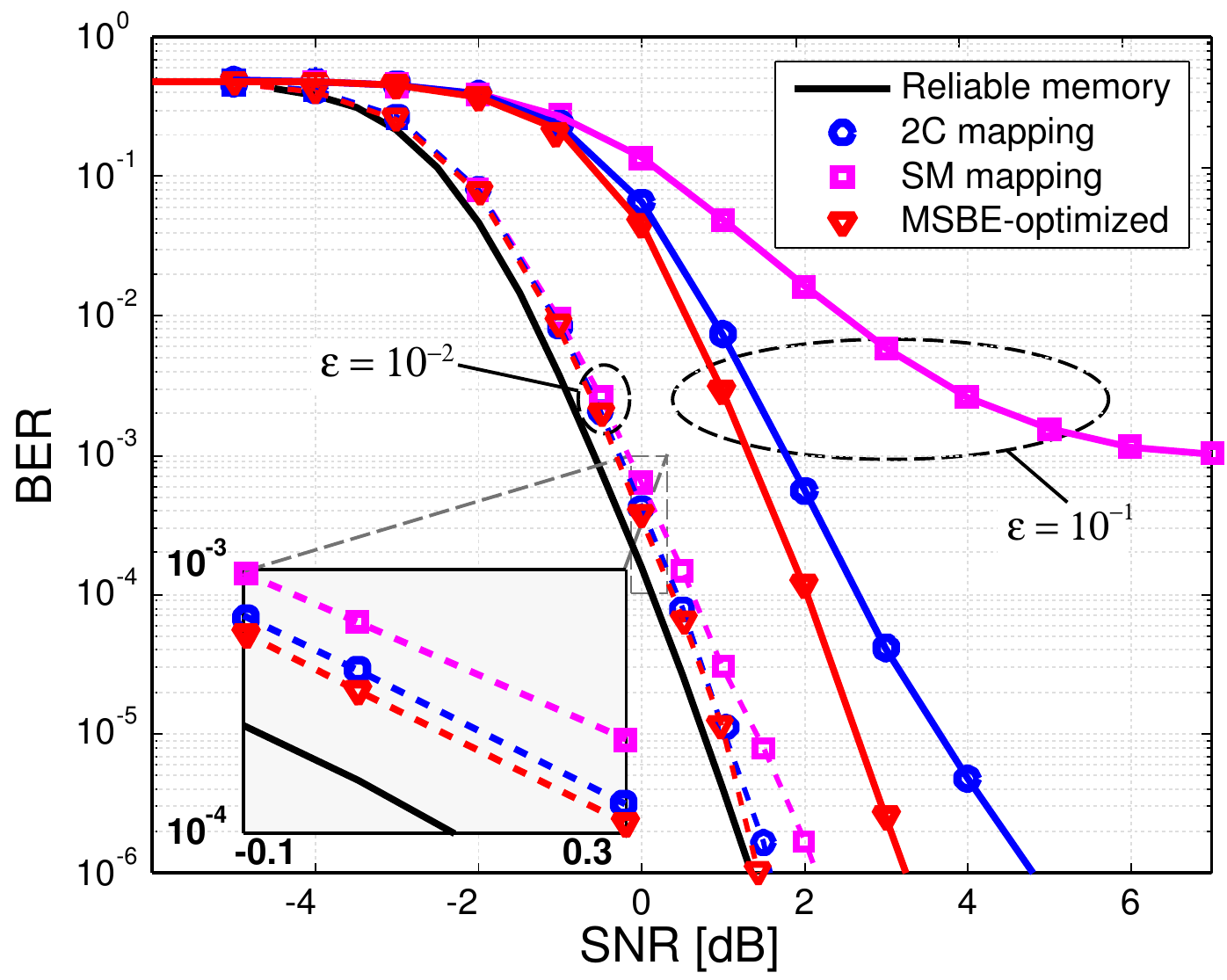}
\figspace
\caption{BER-performance comparison of convolutional coding for
  different data representations. Solid curves represent the
  results for \mbox{$\varepsilon = 10^{-1}$}, dashed curves the results for 
  $\varepsilon = 10^{-2}$. The quantizer parameters are $N=3$, $I=2$, $F=0$, and $\delta=0.6$.}
\label{fig:berVit}
\end{figure}

As a second example, we assume that the encoder in
Fig.~\ref{fig:commSysModel} corresponds to the rate-1/2, 256-state
convolutional code as specified in 3GPP-HSPA~\cite{3GPP_TS25222}. On the receiver side, we deploy a
soft-input Viterbi decoder to generate estimates for the transmitted bits~\cite{viterbi1967}. We note that convolutional codes remain to be the most common channel codes found in today's (wireless) communication systems and, thus, the robustness of the Viterbi decoder against memory defects is of significant practical interest.
In this example, the unreliable LLR memory corresponds to the (typically large) memory used for 
LLR de-interleaving. We are again interested in the effect of data representations on the BER performance for systems employing unreliable memories.

\subsubsection{Cost function for Viterbi decoding}
The derivation of  an analytical expression for the error-rate performance $P_e$
of the soft-input Viterbi decoder is difficult, in general. 
It is, however, important to realize that the Viterbi algorithm~\cite{viterbi1967} computes its decisions on the basis of \emph{distances} between branch metrics, which are used to find the maximum-likelihood solution. These branch metrics correspond to summations of LLR values, and, hence, are directly affected by the unreliable LLR memory. 
Consequently, the
unreliable LLR memory induces branch-metric cross-over probabilities
from branch metrics $\mathrm{bm}$ computed at the input of the
unreliable LLR memory to the branch metrics $\overline{\mathrm{bm}}$
 at its output
\begin{align*}
    P_{\mathrm{\overline{bm}}|{\mathrm{bm}}}(\overline{\mathrm{bm}}_{m'}|\mathrm{bm}_{m}) =
    \mathrm{Pr}(\overline{\mathrm{bm}}= \overline{\mathrm{bm}}_{m'} |
    \mathrm{bm} = \mathrm{bm}_m),
\end{align*}
which are a function of the memory cross-over probability
$P_{\mathrm{C,SAC}}$ and the LLR distribution $P_d$. 
Furthermore, the branch metrics computed at the input of the unreliable memory follow
the distribution $P_{\mathrm{bm}}$, a consequence of $P_d$. 
To arrive at a cost function that preserves distances between the branch metrics, we propose to minimize the mean squared branch-metric error~(MSBE)
\begin{align}\label{eq:msbe}
    \text{MSBE}  = & 
    \sum_{m=1}^M P_{\mathrm{bm}}(\mathrm{bm}_m) \times \notag \\
   & \sum_{m'=1}^M (\mathrm{bm}_m-\overline{\mathrm{bm}}_{m'})^2 P_{\mathrm{\overline{bm}}|{\mathrm{bm}}}(\overline{\mathrm{bm}}_{m'}|\mathrm{bm}_{m}),
\end{align}
i.e., we set the cost function to $\setC_{\text{MSBE}} = \text{MSBE}$.

\subsubsection{Simulation results and discussion}
In Fig.~\ref{fig:berVit}, we show the BER-performance of the system 
for different data representations and two memory bit-cell error probabilities, i.e., $\varepsilon
=10^{-1}$~(solid curves) and $\varepsilon = 10^{-2}$~(dashed 
curves). We set $N=3$ and optimize  the data mapping separately for each
SNR point by exhaustive evaluation of
$\setC_{\text{MSBE}}$. The BER-performance of these optimized
data mappings is shown in Fig.~\ref{fig:berVit} (labeled by `MSBE-optimized').

We observe that the chosen data representation has a significant impact on the performance of the soft-input Viterbi
algorithm. Hence, careful selection of the data mapping function is of paramount importance in systems containing unreliable memories. We furthermore observe that the cost function proposed in~\eqref{eq:msbe} is able to significantly increase the robustness against unreliable memories for communication systems relying on convolutional codes compared to the use of conventional number representations, such as 2C and SM.


\section{Conclusions} 
\label{sec:conclusion}

In this paper, we have studied the impact of unreliable
memories on digital signal processing (DSP) systems. In particular, we have introduced the
\emph{stuck-at-channel} as a realistic model for bit-cells in unreliable memories, such as high-density memories implemented in advanced semiconductor technologies.
We have observed that the data representation must be chosen carefully according to a cost function suitable for the given application to minimize the impact of unreliable memories on the system performance.
To highlight the efficacy of the proposed framework, we have considered  two practical application examples 
in a coded wireless communication system.
For repetition coding and convolutional codes, our results demonstrate that the deployment of optimized data representations enables substantial gains in terms of the error-rate performance. 
In particular, we find that the two most common data representations used in digital integrated systems, namely 2's complement and sign-magnitude, are generally outperformed by data mappings optimized for the application at hand. 
Hence, the proposed data recoding method can be considered to be an effective low-complexity method to improve the robustness of general DSP systems that contain unreliable memories.


%
%

\bibliographystyle{IEEEtran}

\vspace{0.135cm}

{
\bibliography{IEEEabrv,12ALLER_DataMapping.bib}
}


\end{document}